\documentclass[twocolumn,aps,nofootinbib]{revtex4} 
\usepackage{amsfonts, amssymb}
\usepackage{graphicx, epsfig,bm}

\newcommand{\be}{\begin{equation}}
\newcommand{\ee}{\end{equation}}
\newcommand{\bea}{\begin{eqnarray}}
\newcommand{\eea}{\end{eqnarray}}

\newcommand{\ApJ}{{\it Astrophys. J.\,}}
\newcommand{\ApJS}{{\it Astrophys. J. Suppl.\,}}

\newcommand{\etal}{{\it et al.}}

\def\fun#1#2{\lower3.6pt\vbox{\baselineskip0pt\lineskip.9pt
        \ialign{$\mathsurround=0pt#1\hfill##\hfil$\crcr#2\crcr\sim\crcr}}}

\newcommand\Mpc{\,\mbox{Mpc}}

\newcommand\lsim{\mathrel{\rlap{\lower4pt\hbox{\hskip1pt$\sim$}}
    \raise1pt\hbox{$<$}}}
\newcommand\gsim{\mathrel{\rlap{\lower4pt\hbox{\hskip1pt$\sim$}}
    \raise1pt\hbox{$>$}}}

\def\dslash{\not{\hbox{\kern-2pt $\partial$}}}
\def\Dslash{\not{\hbox{\kern-4pt $D$}}}
\def\Oslash{\not{\hbox{\kern-4pt $O$}}}
\def\Qslash{\not{\hbox{\kern-4pt $Q$}}}
\def\pslash{\not{\hbox{\kern-2.3pt $p$}}}
\def\kslash{\not{\hbox{\kern-2.3pt $k$}}}
\def\qslash{\not{\hbox{\kern-2.3pt $q$}}}

 \newtoks\slashfraction
 \slashfraction={.13}
 \def\slash#1{\setbox0\hbox{$ #1 $}
 \setbox0\hbox to \the\slashfraction\wd0{\hss \box0}/\box0 }
 
\def\ee{\end{equation}}
\def\be{\begin{equation}}

\begin{document}
\setlength{\unitlength}{1mm}
%\twocolumn[\hsize\textwidth\columnwidth\hsize\csname@twocolumnfalse\endcsname]
\title{Did Boomerang hit MOND ?}
\author{An\v{z}e Slosar$^1$, Alessandro Melchiorri$^2$, Joseph I. Silk$^3$}
\affiliation{ 
$^1$Faculty of Mathematics and Physics, University of Ljubljana, Slovenia\\
$^2$ Physics Department and sezione INFN, University of Rome ``La Sapienza'', Ple Aldo Moro 2, 00185 Rome, Italy\\
$^3$ Astrophysics, Denys Wilkinson Building, Keble Road, OX13RH, Oxford, United Kingdom\\}
\date{\today}%
%\maketitle % use with old revtex ! 

\begin{abstract}
  
  Purely baryonic dark matter dominated models like MOND based on
  modification of Newtonian gravity have been successfull in
  reproducing some dynamical properties of galaxies. More recently, a
  relativistic formulation of MOND proposed by Bekenstein seems to
  agree with cosmological large scale structure formation.  In this
  work, we revise the agreement of MOND with observations in light of
  the new results on the Cosmic Microwave Anisotropies provided by the
  $2003$ flight of Boomerang.  The measurements of the height of the
  third acoustic peak, provided by several small scale CMB experiments
  have reached enough sensitivity to severely constrain models without
  cold dark matter. Assuming that acoustic peak structure in the CMB
  is unchanged and that local measurements of the Hubble constant can
  be applied, we find that the cold dark matter is strongly favoured with 
  Bayesian probability ratio of about one in two hundred.

\end{abstract}
\bigskip
%\pacs{PACS Numbers: }

\maketitle 

\section{Introduction}

The measurements of the Cosmic Microwave Background (CMB) anisotropies,
most notably by the Wilkinson Microwave Anisotropy Probe (WMAP) mission
\cite{Bennett03}, have truly marked the beginning of the era of
precision cosmology. In particular, the shape of the measured
temperature and polarisation angular power spectra are in spectacular
agreement with the expectations of the standard model of structure
formation, based on primordial adiabatic and nearly scale invariant
perturbations (see e.g. \cite{spergel}). More recently, ground based 
and balloon borne CMB experiments like extended VSA \cite{vsa},  CBI 
\cite{cbi}, Acbar \cite{acbar1}, DASI \cite{dasipol} and BOOMERANG-03 \cite{boom03}, 
have probed the CMB power spectra at smaller scales, 
confirming the presence of acoustic oscillations and 
providing the first unambiguous detection of polarisation.

Moreover, new, complementary, results from the Sloan Digital Sky
Survey (SDSS) on galaxy clustering (see e.g. \cite{Tg04}) and, more
recently, on Lyman-$\alpha$ Forest clouds \cite{Se04} are now further
constraining the scenario.

Since all these measurements appear in spectacular agreement with the
$\Lambda$CDM model, based on a cosmological constant and 
on cold dark matter, is definitely timely to investigate what space
is left for alternative theories. 

Perhaps the most exotic alternative to the standard model
one could consider is MOdified Newtonian Dynamics (MOND, \cite{M1})
where a purely baryonic model with modifications to standard (Newtonian) 
gravity is suggested. In MOND the departure from Newtonian law
$\mathbf{a} =-\bm{\nabla}\Phi_{\rm N}$ is given by:  
\begin{equation}
\tilde\mu(|\mathbf{a}|/\mathfrak{a}_0)\mathbf{a} =
-\bm{\nabla}\Phi_{\rm N} 
\label{MOND}
\end{equation} where $\Phi_{\rm N}$ is the Newtonian potential
of the visible matter, $\mathfrak{a}_0$ is an acceleration scale 
while $\tilde\mu(x)\approx x$ for
$x\ll 1$ and 
$\tilde\mu(x)\rightarrow 1$ for
$x\gg 1$. If $\mathfrak{a}_0\approx 1\times
10^{-8}$ cm s$^{-2}$ the Newtonian law is recovered in the solar 
system where accelerations are large compared to
$\mathfrak{a}_0$.

The above empirical formula (\ref{MOND}) has been originally proposed
to explain the fact that rotation curves of disk galaxies become flat
outside their central parts.  While in the standard dark matter
paradigm flat rotation curves are explained by assuming a spherical
halo of invisible dark matter around visible disk galaxies, in MOND
there is no need to include non baryonic dark matter since, thanks to
the Eq.\ref{MOND}, galaxies far out exhibit an approximately spherical
Newtonian potential without the inclusion of the dark matter. Attempts
were made to confront MOND with clusters of galaxies (see
e.g. \cite{sand03,poi05}) and the large scale structure
(e.g. \cite{adi}) with mixed success.
 
While the non-baryonic dark matter paradigm is definitely more
compelling for its aesthetic simplicity than a modification to
Newtonian gravity, the MOND model has been claimed successful in
other aspects (see e.g. \cite{mcg98,san2k}) and MOND proponents
insist that this alternative model for gravity merits serious
examination.

The MOND theory has suffered from a lack of a successful relativistic
formulation, that would allow one to compare it to observations of CMB
and Large Scale Structure.  Nevertheless, there were some attempts to
confront CMB data \cite{gaugh04}, which find that the first $2$ observed 
acoustic peaks in the CMB spectrum are compatible with MOND at the 
price of a substantial neutrino mass,
which is barely compatible with current laboratory bounds \cite{fogli},
or at the price of including curvature \cite{lmg}, which is
at odds with the inflationary scenario.

A major step in the direction of developing the simple MOND formula
into a more robust theory of gravity has been recently proposed by
Bekenstein \cite{bekenstein}. In this paper, a relativistic gravitational
theory has been presented whose nonrelativistic weak acceleration
limit accords with MOND while its nonrelativistic strong acceleration
regime is Newtonian.

Moreover, Bekenstein's model provides a specific formalism for
constructing cosmological models and testing MOND using cosmological
data. Indeed, more recently, Skordis et al. \cite{skordis} produced
the first theoretical prediction for CMB anisotropies and Large Scale
Structure in the case of Bekenstein's model. It has been shown that
the Bekenstein model may be put in agreement with the WMAP data and
Large Scale Structure observations.  Similar to previous results
authors find agreement if neutrinos ensure that peak positions are
unchanged.  The results are obviously of great relevance since the
model has no cold dark matter in it and may therefore be considered as
an important alternative to the present cosmological scenario, which
assumes a fine-tuned cosmological constant and yet to be discovered
dark matter particles.

In this brief report we point out that recent, small scale, CMB data
already provide discriminating power between these two scenarios.  In
the standard cold dark matter scenario, the amplitude of the CMB peaks
is sensitive to the amount of dark matter because of two effects:
increasing the matter density on one hand decreases the radiation
driving while on the other hand it increases the depth of potential
wells. These two effects nearly cancel out in the amplitude of the
second acoustic peak, but conspire to produce a higher amplitude of
the third peak (see \cite{hufu}). The height of the second peak to the
first peak therefore contains information on the baryonic content of
the Universe, while the ratio of the third peak to the first peak
height tells us about the matter density.

A generic prediction of purely baryonic dominated models like MOND is
therefore that peaks in the CMB power spectrum should be strictly
decreasing in amplitude.

Extraordinary WMAP results on the first two peaks, coupled with the
recent small scale CMB data on the third peak now have enough power to
discriminate between these two scenarios and to determine the amount
of cold dark matter.

The goal of this brief report is to examine the data on the third peak, with
special emphasis on the recent measurements of the CMB fluctuations by
the Boomerang experiment.

The ultimate test of MOND, would be a full confrontation of the
relativistic theory with the data. This is a daunting task, given that
the theory is very complicated with yet to be fully understood
perturbation theory and several free parameters including a free
function. We note, however, that models discussed in \cite{skordis}
are in a complete agreement with generic $\Lambda$-CDM predictions in
the range $\ell>200$. We therefore assume that this is a generic
prediction of the Bekenstein's theory and proceed by fitting the
$\ell>200$ region of the CMB data with the standard $\Lambda$-CDM
models to see whether models with zero cold dark matter density are
compatible with the data. Whether our assumption is a justified one is
to be seen, however, we feel it nevertheless provides a first order
confrontation of the data with the theory. Our approach is orthogonal to
that of \cite{skordis} in the sense that it provides constraints on any theory
that leaves the CMB physics  unchanged on scales smaller or  roughly
equal to that of the first acoustic peak.

\medskip
\section{Analysis}
\medskip

We use the \texttt{Cosmo-MC} package (\cite{cosmomc}) to perform parameter
estimation on standard flat $\Lambda$-CDM models using top-hat priors
on the following 6 parameters: $\omega_{\rm b}=\Omega_{\rm b} h^2$
(the baryon density of the universe), $\omega_{\rm dm}$ (the dark
matter density of the universe), $\theta$ (ratio of the sound horizon
to the angular diameter distance to the surface of last scattering,
multiplied by 100), $\nu_{\rm frac}$ (the fraction of dark matter in
form of massive neutrinos), $n_{\rm s}$ (spectral index of primordial
fluctuations), $\log A_{\rm s}$ (the amplitude of primordial scalar
fluctuations). The priors used are listed in the Table
\ref{tab:prior}.

\begin{table}
  \centering
  \begin{tabular}{cc}
  Parameter & Prior\\
\hline 
$\omega_{\rm b}$ & BBN (see text) \\
$\omega_{\rm dm}$ & (0,0.99) \\
$\theta$ & (0.5,10) \\
$\nu_{\rm frac}$ & (0,1) \\
$n_{\rm s}$ & (0.5,1.5) \\
$\log 10^{10} A_{\rm s}$ & (2,5) \\
 $n_{\rm run}$ & (-0.2,0.2)
  \end{tabular}
  \caption{Flat priors on the cosmological parameters. Notation
  $(x,y)$ implies a a flat prior between $x$ and $y$. Running spectral
  index (last parameter) was used only in the part discussed in the
  last paragraph of section \ref{sec:results}.}
  \label{tab:prior}
\end{table}

In our parametrisation the density of the \emph{cold}  dark matter is
given by 
\begin{equation}
  \Omega_{\rm cdm} = \frac{\omega_{\rm dm}\left(1-\nu_{\rm frac}\right)}{h^2}
\end{equation}

We intentionally omitted $\tau$, the optical depth to the last
scattering, from our parametrisation as it is completely degenerate
with the amplitude in the multipoles of interest. 

The following datasets were used in our parametrisation: WMAP
\cite{Bennett03,verde}, VSA \cite{vsa}, CBI \cite{cbi}, Acbar
\cite{acbar1}, and the latest Boomerang results \cite{boom03}.  In
these datasets we have used the default \texttt{Cosmo-MC} distribution
datasets for VSA, CBI and Acbar experiments. In all datasets any
points with $\ell < 200$ were removed and additionally all points with
$\ell<375$ were removed from Boomerang temperature data (to prevent
cosmic variance coupling to the WMAP dataset).

The basis of our analysis are the heavily oversampled chains of the
WMAP dataset (with about $\sim$260,000 independent points) on the
parametrisation described above, together with a Big Bang
Nucleosynthesis (BBN) prior of $\omega_b=0.020\pm0.002$ \cite{bbn}. There
is an additional prior on $0.4<h<1.0$ hard-coded into the
\texttt{Cosmo-MC} code and our posterior space is cut by the lower end
of this prior. Releasing this prior would weaken our constraints, but
$h<0.4$ models would face many difficulties with other cosmological probes.

  We importance sample these chains with additional data in order to
get improved constraints. Apart from savings in the CPU time, the
importance sampled chains still include models from the low
$\Omega_{\rm dm}$ region which is of interest for the MOND models and
would otherwise be present only far into the tails of the probability
distribution. We also made a similar run without the BBN prior but
with very wide flat prior on $\omega_{\rm b}$ instead, to check the
effect of BBN prior dependence.

\medskip
\section{Results}
\medskip
\label{sec:results}

\begin{figure}[t]
\begin{center}
\includegraphics[height=\linewidth, angle=-90]{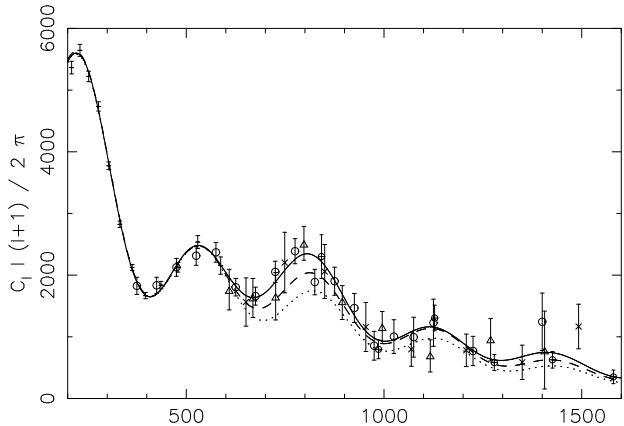}
\caption{This plot shows the experimental data used and a few
  theoretical spectra. Points with large error-bars were removed from
  the plot (but actually used in the parameter estimation chains) to
  maintain clarity.  The experiments are as follows: WMAP (bars), VSA
  (triangles), CBI (crosses), ACBAR (crossed circles), Boomerang
  (circles). The theoretical graphs plotted are most likely model in
  our chains for the all CMB data (solid line) and two models picked
  from best fit models with $\Omega_m<0.01$ from each chain sampling
  WMAP data only. Parameters for dashed model are $(\omega_b,
  \omega_\nu,h,n_s)=(0.22,0.096,0.76,0.59)$ and those for the dotted
  are $(0.22,0.15,0.78,0.43)$.  See text for discussion.  \label{data-plot}}
\end{center}
\end{figure}

The parameter of major interest here is $\Omega_{\rm cdm}$. A MOND
predicts a zero $\Omega_{\rm cdm}$, while the standard prediction is
that of a $\Omega_{\rm cdm}$ of about 0.3. In the Figure
\ref{data-plot} we plot the data we used (note that some points were
omitted from the plot - see caption) and a few theoretical
predictions.
 The solid line model corresponds to a standard $\Lambda$CDM flat
model and fits the data very well. The dashed and dotted models were
cherry-picked from the models that have the highest likelihood of
models that satisfy $\Omega_m<0.01$ in each individual MCMC chain
(using WMAP data alone). The dashed model illustrates the common wisdom
that it is possible to construct models that have nearly identical
peak positions and heights of even peaks with zero dark
matter. Finally, the dotted model shows an example of a zero dark
matter model that is allowed by the WMAP data but obviously at odds
with measurements of the small scale power.

\begin{figure}[t]
\begin{center}
\includegraphics[height=\linewidth, angle=-90]{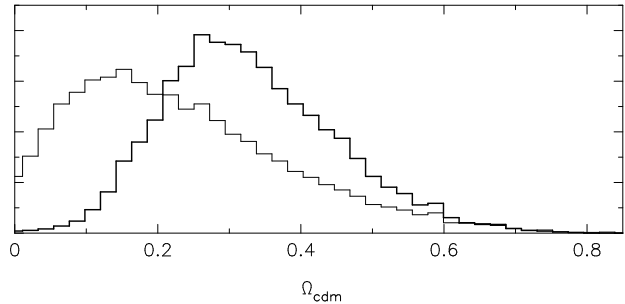}
\caption{This plot shows the marginalised probability distribution for
  $\Omega_{\rm cmd}$ for the WMAP dataset alone (thin line) and
  WMAP/VSA/ACBAR/CBI/Boomerang datasets (thick line). Curves are
  normalised to the same total area under the curve.\label{fig:1}}
\end{center}
\end{figure}

 In the Figure \ref{fig:1} we plot the marginalised probability
distribution for $\Omega_{\rm cdm}$ for the WMAP dataset and the WMAP
dataset after inclusion of all the other CMB data. We note that the
WMAP alone admits the zero cold dark matter solutions, in agreement
with previous investigations \cite{skordis,gaugh04}. The
addition of other datasets, however, strongly rejects this region of
the parameter space, without resorting to non-CMB experiments.

In order to quantify this further, we use two statistical tests.
Firstly, we compare two basic models, namely the flat CDM model with
$\Omega_{\rm cdm}$ between 0 and 1 and the MOND model for which
$\Omega_{\rm cdm}=0$ using Bayesian model comparison \cite{slal}. In
order to do this we compute the logarithm of evidence ratio with a
version of the Savage-Dickey test, equivalent to that described in the
Appendix of \cite{ss04}. Secondly we estimate the number of ``sigmas''
in a frequentist manner by comparing the likelihoods of the
$\Omega_{\rm cdm}=0$ point and the most likely point in the
marginalised probability distribution for $\Omega_{\rm cdm}$. The
likelihood ratio can be converted to a number of standard deviations
using the prescription $ n_{\sigma}=\sqrt{2\frac{\mathcal{L}_{\rm
      max}}{\mathcal{L}_{\rm MOND}}}$, which returns the expected
result in the Gaussian case. In such short communication it is
impossible to compare the two methods in depth, but we note that
frequenstist approaches neary always give higher ``detection''
confidences. Bayesian evidence ratio has an advantage that it directly
encodes the probabiltiy ratio.

The results are summarised in the Table \ref{tab:res}. The exact
numbers somewhat depend on the binning width and therefore the numbers
in the table are accurate to about $0.1$ in both columns. Some row
state the constraints upon adding two extra constraints on top
of all CMB data. HST label denotes the Hubble Space Telescope (HST)
constraint on the Hubble's constant \cite{freedman}, which is
conveniently described as a Gaussian around $h=H_0/100 {\rm
  km/s/Mpc}=.72$ with 1-$\sigma$ dispersion of $0.8$. The HST data
actually has a rather strong effect on our results as the Hubble
constant favoured by the $\Omega_{\rm cdm}=0$ models is rather low. SN
label denotes the gold dataset of the \cite{riess} supernovae data.
These results must be taken with some caution, because it is not
entirely clear that the standard interpretation of these two
cosmological probes is applicable in the MONDian setting.

\begin{table}
  \centering
  \begin{tabular}{p{5.5cm}p{0.5cm}cc}
    Dataset &  & $\Delta \log E$ & $n_\sigma$ \\
\hline
    WMAP                         & &      0.1   &       1.5 \\
    WMAP/VSA/ACBAR/CBI     & &      2.1   &        2.6 \\
    WMAP/VSA/ACBAR/CBI/BOOM  & & 3.6  &       3.1 \\
\hline
     +HST                  & &    5.2  &        3.6 \\
     +HST +SN               & &    10.5 &        5.1 \\
\hline
    ALL CMB w/o BBN        & & 2.2 &  2.6 \\
    -BBN +HST             & & 3.2 & 3.3 \\
    -BBN +HST +SN         & & 4.3 & 3.6 \\
  \end{tabular}
  \caption{This table shows results of the two statistical tests
    described in the text for a range of datasets considered. The sign
    convention is such that higher number implies higher statistical
    evidence in favour of cold dark matter models. The $\Delta \log E$
    is trivially interpreted as the logarithm of probability
    ratio. See text for discussion.}
  \label{tab:res}
\end{table}

Finally, we have also considered running of the spectral index,
defined by $n_{\rm run}={\rm d } n_{\rm s}/{\rm d} \ln k$ with pivot
scale set to $k=0.05\, \Mpc\,{h}^{-1}$.  Using a top-hat prior between
-0.2 and 0.2 on $n_{\rm run}$ we find that this parameter is
completely unconstrained by the WMAP data and only weakly constrained
by the all CMB data. However, the very high $\ell$ points from Acbar seem
to favour \emph{negative} running indices, thus even lowering the
third peak required by MOND.  Consequently we find that the bound on
$\Omega_{\rm cdm}$ is unaffected when all CMB data are included.

\medskip
\section{Discussion and Conclusions}
\medskip

In this brief report we have analysed the latest CMB data in light of
the recently renewed interest in the MOND theories of gravity. We
simplified (and potentially oversimplified) the theory by assuming
that the small scale CMB fluctuations are unmodified by the MOND
theory in accordance with recent attempt \cite{skordis} to model
linear fluctuations in the relativistic theory of MOND recently
proposed by Bekenstein. Under this assumption we find that the WMAP
data alone is fully consistent with the $\Omega_{\rm dm}=0$ required
by MOND, in accordance with previous findings.  A component of massive
sterile neutrinos is required with $\omega_\nu \gtrsim 0.1$, which is
marginally consistent with earth-based beta electron decay
experiments.  Addition of other CMB data constrains the third peak
height, which encodes the information on the presence of the cold dark
matter. The data before the latest Boomerang dataset weakly favour the
cold dark matter but only at 1 in 8 probability ratio for a Bayesian
analysis and less than 3-$\sigma$ for the more conventional approach.
The VSA's measurement of power at the top of the third peak is
probably crucial in absence of Boomerang data. Addition of the third
peak from latest Boomerang data gives crucial extra information.  The
Bayesian probability ratio in favour of cold dark matter models
increases to 1 in 36, while the likelihood ratio breaks the 3-$\sigma$
``barrier''. We note that using less general models for the CDM
setting would make the Bayesian model selection even stronger in
favour of cold dark matter models.

Releasing the BBN prior somewhat weakens the constraints. We note
however, that there is no obvious mechanism how could a MOND theory
evade BBN constraints and that such models are additionally
characterised by very low values of $h$ and $n_s$. Marginalised value
of $\omega_{b}\sim 0.025$ for these models.

Finally, we have added two other standard cosmological probes, the HST
measurements of the Hubble's constant and the recent supernovae
measurements of the luminosity distance. Taken at the face value, they
seem to blow the MOND model into oblivion. Care must be taken,
however, in interpreting these two datasets as it is not clear whether
it is appropriate to include them in the MONDian scenarios without
detailed treatment of possible MONDian effects on the background
evolution. 

We have shown that running of the spectral index cannot rescue the
third peak, at least for $|n_{\rm run}|<0.2$. 

We note that our results are somewhat prior dependent: the
$\Omega_{\rm cdm}$ parameter is a derived parameter in the standard
parametrisation used by the \texttt{Cosmo-MC} package and therefore
the implied prior on it is certainly not flat. However, we feel that
priors employed are actually a sound set of physical priors and
therefore our results should be fairly insensitive to any sensible
reparametrisations.

Finally, we note that there is still a possibility that a version of
MOND theory with a high third peak is constructed. However, this would
seem rather artificial. The ratio of height of third to the first peak
encodes the amount of the baryonic drag and in order to get a high
third peak one needs some cold dark matter like element. Even if this
eventually turns out to be a scalar field in a MONDian theory, the
present data indicate that dark matter theory is at least a very good
approximation to the full underlying theory.

\textit{Acknowledgements} We thank Antony Lewis for kindly providing
us with a version of \texttt{Cosmo-MC} supporting the new
\texttt{.newdat} format before its official release. AS acknowledges
useful discussion with Uro\v{s} Seljak. 
We acknowledge the use of the COSMOS cluster (an UK-CCC facility)

%% AS acknowledges support from
%% the Slovene Ministry of Science, Education and Sport. AM is supported
%% by MURST through COFIN contract no. 2004027755. We acknowledge the use
%% of the COSMOS cluster and thank the sponsors of this UK-CCC facility,
%% supported by HEFCE and PPARC. 
%% This research was conducted in
%% cooperation with GSI/Intel utilising the Altix 3700 supercomputer.

\end{document}